\documentclass[11pt]{article}
\usepackage{epsfig}
\usepackage{amsfonts}
\usepackage{amscd}
\usepackage{latexsym}
\usepackage{amsmath,amssymb}
\usepackage{verbatim}
\usepackage{setspace}
\usepackage{color}
\usepackage{fancyhdr}
\usepackage{cite}
\usepackage{hyperref}
\usepackage{tensor}
\usepackage{xfrac}

\usepackage[textheight=9in, textwidth=6.5in, letterpaper]{geometry}
\def\half{{1\over 2}}
\numberwithin{equation}{section}

\def\e{{\epsilon}}

 \def\p{\partial}
 \def\bz{{\bar z}}
 
\def\0{{(0)}}
\def\1{{(1)}}
\def\2{{(2)}}
 \def\cL{{\cal L}}

\def\<{\langle }
\def\>{\rangle }

\def \lam {\lambda}
\def \eps {\epsilon}
\def\bz{{\bar z}} \def\zb{{\bar z}}
\def \e {\epsilon}

\def\p{\partial}

\def\CL {{\cal L}}

\def\CQ {{\cal Q }}

\def\csm {{\cal C \cal S^- }}
\def\csp {{\cal C \cal S^+ }}

\def\zb {{\bar{z}}}

\newcommand{\bea}{\begin{eqnarray}}
\newcommand{\eea}{\end{eqnarray}}
\newcommand{\be}{\begin{equation}}
\newcommand{\ee}{\end{equation}}
\newcommand{\ba}{\begin{aligned}}
\newcommand{\ea}{\end{aligned}}

\def\be{\begin{equation}}
\def\ee{\end{equation}}
\def\beq{\be\begin{array}{c}}
\def\eeq{\end{array}\ee}

\renewcommand{\epsilon}{\varepsilon}

   \makeatletter
  \let\over=\@@over \let\overwithdelims=\@@overwithdelims
  \let\atop=\@@atop \let\atopwithdelims=\@@atopwithdelims
  \let\above=\@@above \let\abovewithdelims=\@@abovewithdelims
\renewcommand\section{\@startsection {section}{1}{\z@}%
                                   {-3.5ex \@plus -1ex \@minus -.2ex}
                                   {2.3ex \@plus.2ex}%
                                   {\normalfont\large\bfseries}}

\renewcommand\subsection{\@startsection{subsection}{2}{\z@}%
                                     {-3.25ex\@plus -1ex \@minus -.2ex}%
                                     {1.5ex \@plus .2ex}%
                                     {\normalfont\bfseries}}

\usepackage{tikz}
\usepackage{slashed}
\usetikzlibrary{calc}   
\usetikzlibrary{topaths}
\usetikzlibrary{decorations}
\usetikzlibrary{decorations.pathmorphing}

{\cfoot{\thepage}}
\begin{document}
\begin{titlepage}
\unitlength = 1mm~\\
\vskip 3cm
\begin{center}

{\LARGE{\textsc{Uplifting AdS$_3$/CFT$_2$ to Flat Space Holography}}}

\vspace{0.8cm}
Adam Ball, Elizabeth Himwich, Sruthi A. Narayanan, Sabrina Pasterski, and Andrew Strominger
\vspace{1cm}

{\it  Center for the Fundamental Laws of Nature, Harvard University,\\
Cambridge, MA 02138, USA}

\vspace{0.8cm}

\begin{abstract}
Four-dimensional (4D) flat Minkowski space  admits a foliation by hyperbolic slices. Euclidean AdS$_3$ slices fill the past and 
future lightcones of the origin, while dS$_3$ slices fill the region outside the lightcone. The resulting link between 4D asymptotically flat quantum gravity  and AdS$_3$/CFT$_2$ is explored in this paper.  The 4D superrotations in the extended BMS$_4$ group are found to act as the familiar conformal transformations on the 3D hyperbolic slices, mapping each slice to itself. The associated 4D  superrotation charge is constructed in the covariant phase space formalism. The soft part gives the 2D stress tensor, which acts on the celestial sphere at the boundary of the hyperbolic slices, and is shown to be an uplift to 4D of the familiar 3D holographic AdS$_3$ stress tensor. Finally, we find that 4D quantum gravity  contains an unexpected  second, conformally soft, dimension $(2,0)$ mode that is symplectically paired with the celestial stress tensor.
\end{abstract}

\end{center}

\end{titlepage}

\tableofcontents

\section{Introduction}
\begin{figure}[htb]
\center
\includegraphics[scale=4.0]{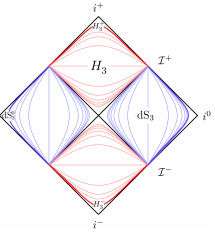}
\caption{Penrose diagram of hyperbolic slicing of Minkowski space. The slices correspond to surfaces of constant $\tau$. The slices in the past and future lightcones of the origin have the geometry of $H_3$, and the slices with spacelike separation from the origin have the geometry of dS$_3$. }
\label{slicing}
\end{figure}

The metric for flat 4D  Minkowski space ($M_4$) in hyperbolic coordinates is
\begin{equation}
\label{eq:flatone}
ds^2=-d\tau^2+\tau^2\left(\frac{d\rho^2}{\rho^2}+\rho^2 dzd\bz\right)
\end{equation}
where $\tau$ is the Lorentz-invariant  distance from the origin and labels the three-dimensional hyperbolic slices in the parenthesis. In order to cover all of $M_4$ we take $\tau$ positive in the future lightcone of the origin, negative in the past lightcone and both $\tau$ and $\rho$ imaginary outside the origin; see Figure~\ref{slicing}. Equation~\eqref{eq:flatone} represents $M_4$ as a kind of non-compact compactification to AdS$_3$. Hyperbolic slicings have been studied for example in~\cite{deBoer:2003vf,Campiglia:2015qka,Campiglia:2015kxa,Cheung:2016iub}.\footnote{See e.g.~\cite{Gary:2009ae,Fitzpatrick:2011jn} for an alternate approach to  $M_4$ holography as the flat space limit of AdS$_4$ quantum gravity rather than an uplift of AdS$_3$ quantum gravity. }

In this paper, we take inspiration from the prescient paper of de Boer and Solodukhin~\cite{deBoer:2003vf}. These authors conjectured that the infinite-dimensional 2D conformal symmetry of AdS$_3$ quantum gravity should uplift to $M_4$ quantum gravity, with separate symmetries  for the past and the future. Somewhat later, the existence of such conformal symmetries, coined superrotations, was conjectured in~\cite{Barnich:2010eb,Barnich:2011ct,Barnich:2011mi,Barnich:2016lyg} by relaxing an overly-restrictive assumption about the asymptotic behavior of the gravitational field in the original papers of BMS~\cite{Bondi:1960jsa,Bondi:1962px,Sachs:1962zza}. More recently~\cite{Kapec:2014opa,Kapec:2016jld}, using the subleading soft theorem of~\cite{Cachazo:2014fwa}, the existence of a single conformal symmetry of quantum gravitational scattering in $M_4$ was proved. The  past-future pair of conformal symmetries of~\cite{deBoer:2003vf,Barnich:2010eb,Barnich:2011ct,Barnich:2011mi,Barnich:2016lyg} was reduced to a single conformal symmetry by a matching condition required for the consistency of the scattering amplitudes. The reduced symmetry acts in the standard fashion on the celestial sphere at null infinity. This  suggests a holographic relation between quantum gravity on $M_4$ and an as-yet-to-be-understood  ``celestial conformal field theory" on the celestial sphere at the boundary.  

Despite the natural role played by the hyperbolic slicing~\eqref{eq:flatone}, much of the work on superrotations  has used retarded Bondi coordinates (see~\cite{Campiglia:2014yka:2015lxa,Campiglia:2015kxa,Cheung:2016iub} for important exceptions). The main reason for this is simply  that research  on asymptotic structure near null infinity over the last half century primarily uses Bondi coordinates and many formulae are readily available; some references are~\cite{Ashtekar:1981bq,Dray:1984rfa,Barnich:2010eb,Barnich:2011ct,Barnich:2011mi,Strominger:2013jfa,Barnich:2016lyg}. However, even the global $SL(2,\mathbb{C})_{\rm Lorentz}$ subgroup is obscure in these coordinates which are not well-suited for the study of superrotations. A central purpose of this paper is to recast some of the recent results into hyperbolic coordinates and elucidate the connection between $M_4$ and  AdS$_3$ holography.  One hopes that our detailed understanding of AdS holography can be uplifted and applied to flat space holography. 

In Section 2 we present formulae and conventions for the hyperbolic foliation of $M_4$. In Section 3 we show that  superrotations have a simple description in terms of vector fields that are tangent to the slices. In Section 4 we evaluate the boundary and bulk superrotation charges in the covariant phase space formalism. For the bulk expressions, both the soft parts (which are linear in the metric field) and the hard parts (which involve radiation flux) are evaluated as integrals over hyperbolic slices which hug null infinity where the weak field expansion becomes exact. The soft charges are constructed from uplifts of the holographic stress tensor of AdS$_3$ quantum gravity~\cite{Balasubramanian:1999re}, providing a precise relation between $M_4$ and AdS$_3$ holography.  In Section 5 we explicitly evaluate the hard charge for matter sourced by point particles, and find that it reduces to an integral of the subleading soft factor~\cite{Cachazo:2014fwa}. Section 6 demonstrates that the total charge conservation, which involves contributions from two $H_3$ slices and one dS$_3$ slice,  is equivalent to the subleading soft theorem. In Section 7 we relate the soft covariant charges to the celestial stress tensor.  Section 8  identifies a weight $(2,0)$ mode which is not pure gauge and has a canonical symplectic pairing with the superrotation Goldstone mode. This new $(2,0)$ mode is potentially related to new conformally soft theorems and symmetries, but further investigations are left to future work.   The appendix gives details of the linearized Einstein equation in the  hyperbolic slicing.

\section{Preliminaries}
In  hyperbolic coordinates $(\tau,\rho, z,\bz)$
 the Minkowski metric takes the form 
\begin{equation}
\label{eq:flat}
ds^2=-d\tau^2+\tau^2\left(\frac{d\rho^2}{\rho^2}+\rho^2 dzd\bz\right).
\end{equation}
These are related to the usual Cartesian coordinates 
\begin{equation}
ds^2=-(dX^0)^2+(dX^1)^2+(dX^2)^2+(dX^3)^2 
\end{equation}
by
\begin{eqnarray}
\label{hyperboliccoord}
\tau&=&\sqrt{(X^0)^2-(X^1)^2-(X^2)^2-(X^3)^2} \cr
z&=&\frac{X^1+iX^2}{X^0+X^3}\cr
\rho&=&{X^0+X^3 \over \sqrt{(X^0)^2-(X^1)^2-(X^2)^2-(X^3)^2}},
\end{eqnarray}
with inverse 
\begin{eqnarray}
X^0&=&\frac{1}{2}\tau\rho(1+z\bar{z}+\rho^{-2})\cr
X^1&=&\frac{1}{2}\tau\rho(z+\bz)\cr
X^2&=&-\frac{i}{2}\tau\rho(z-\bz)\cr
X^3&=&\frac{1}{2}\tau\rho(1-z\bar{z}-\rho^{-2}).
\end{eqnarray}
 The hyperbolic coordinates represent Minkowski spacetime as a foliation (labelled by $\tau$) of 3D constant curvature hyperbolic  spaces. 
We label the  spacelike  slices in the future  (past) lightcone of the origin by $\tau>0$ ($\tau<0$). We are especially interested in the 
$\pm \tau\to\infty$ slices which approach $\mathcal{I}^\pm$. We denote them by $H_3^\pm$. The de Sitter slices at spacelike separations from  the origin are labelled by positive imaginary $\tau$. The asymptotic $\tau \to i \infty$ slice is denoted  dS$^0_3$. This is illustrated in Figure~\ref{slicing}. The $\rho=\infty$ boundary of $H_3^+$ will be referred to as the ``future celestial sphere" and denoted $\csp$. The analogously defined past celestial sphere will be denoted $\csm$. 

The nonzero connection coefficients are 
\begin{eqnarray}
\Gamma^\tau_{\rho\rho} = \frac{\tau}{\rho^2}, \ \ \Gamma^\tau_{z\bar{z}} = \frac{\rho^2\tau}{2}, \ \ \Gamma^\rho_{\rho\tau} = \frac{1}{\tau}, \ \ 
\Gamma^{\rho}_{\rho\rho} = -\frac{1}{\rho}\cr \ \ \Gamma^\rho_{z\bar{z}} = -\frac{\rho^3}{2}, \ \ \Gamma^z_{z\tau} = \frac{1}{\tau}, \ \ \Gamma^z_{z\rho} = \frac{1}{\rho}, \ \ \ \Gamma^{\bar{z}}_{\bar{z}\tau} = \frac{1}{\tau}, \ \ \ \Gamma^{\bar{z}}_{\bar{z}\rho} = \frac{1}{\rho}.
\end{eqnarray}

\section{Superrotation Vector Fields}
3D Euclidean quantum gravity on an asymptotically hyperbolic space $H_3$ has a  conformal symmetry which acts  as~\cite{brown1986,Balasubramanian:1999re} 
\begin{equation}
\label{shortvector} 
\zeta_Y = Y^z\p_z - \half \p_zY^z\rho\p_\rho-{1 \over 2 \rho^2}\p_z^2Y^z\p_\bz,
\end{equation}
where $Y^z$ is a conformal Killing vector. This is the conformal symmetry of the holographically dual CFT$_2$ which lives on the $S^2$ boundary~\cite{Maldacena:1997re,Aharony:1999ti}. This 3D vector field lifts to 4D, where it maps the hyperbolic slices to themselves and generates the superrotations of 4D quantum gravity in  asymptotically flat space~\cite{deBoer:2003vf, Barnich:2011ct, Cachazo:2014fwa, Kapec:2014opa}. In hyperbolic coordinates only one component of the 4D metric~\eqref{eq:flat} is  transformed:
\begin{equation}
\label{yex} 
\cL_Yg_{zz} = -{\tau^2 \over 2}\p_z^3Y^z. 
\end{equation}
This term is independent of $\rho$ and therefore sub-subleading in the large $\rho$ expansion of the metric. In the 3D case, this component of the metric is proportional to the holographic 2D stress tensor in the Fefferman-Graham construction~\cite{AST_1985__S131__95_0, Balasubramanian:1999re,2007arXiv0710.0919F}.

A special role will be played in the following by the choice  of vector field
\be \label{yzw}Y^z= {1 \over w-z}.\ee
We define 
\begin{equation}
\label{zy} 
\zeta_w\equiv\zeta_{Y=\frac{1}{w-z}}={1 \over w-z}\p_z - {1 \over 2(w-z)^2}\rho\p_\rho-{1 \over  \rho^2(w-z)^3}\p_\bz.
\end{equation}
Any more general superrotation vector field $\zeta_Y$ can then easily be obtained from $\zeta_w$ via the relation
\begin{equation}
\zeta_Y(z)={1 \over 2\pi i}\oint dw Y^w \zeta_w(z).
\end{equation}

\section{Covariant Phase Space Charge}
In this section we  compute the covariant phase space charge $\CQ^+(\zeta_Y)$ as developed in a number of references including~\cite{Zuckerman:1989cx,1987thyg.book..676C, Lee:1990nz, Iyer:1994ys, Iyer:1995kg, Wald:1999wa, Barnich:2001jy,Avery:2015rga}.

\subsection{Boundary Charge}
Under suitable conditions, the charge $\mathcal{Q}^+(\zeta_Y)$ generates (via Dirac brackets or commutators) the superrotations on spacelike surfaces ending at the future celestial sphere $\csp$. For simplicity we will restrict to situations in which the Bondi news vanishes on  $\csp$.\footnote{A time translation can always be used to position the two-sphere $\csp$ at early times before any news has emerged on ${\cal I}^+$. On the other hand, primaries in a conformal basis~\cite{Pasterski:2016qvg} typically have divergences in the radiation flux at $\csp$~\cite{Donnay:2018neh}.  Our analysis would require modifcations to handle such cases, including additions to the charge as discussed in~\cite{Wald:1999wa}.}  The charge  is given by the formula in e.g.~\cite{Iyer:1994ys,Iyer:1995kg} 
\begin{equation}
\label{iw}
\CQ^+=-{1 \over 16\pi}\int_{\csp}*F =\lim_{\rho\to \infty}{1 \over 32\pi}\int d^2z \rho^3\tau F_{\tau \rho},
\end{equation}
where 
\begin{equation}
F_{\mu\nu} = \frac{1}{2}\nabla_\mu\zeta_\nu h+\nabla_\mu h_\nu{}^\lambda \zeta_\lambda + \nabla_\lambda\zeta_\mu h_\nu{}^\lambda+\nabla_\lambda h_\mu{}^\lambda\zeta_\nu+\nabla_\nu h\zeta_\mu-(\mu\leftrightarrow \nu) ,
\end{equation}
with   $\mu,\nu=0,1,2,3$. Here $h_{\mu\nu}$ denotes the linearized, on-shell  metric perturbations
\begin{equation}
g_{\mu\nu}=\eta_{\mu\nu}+h_{\mu\nu}
\end{equation}
where $\eta_{\mu\nu}$ is given in~\eqref{eq:flat}. Before proceeding further, in order to avoid long expressions, we make the radial gauge choice
\begin{equation}
\label{rg}h_{\tau \mu}=0,
\end{equation}
which can also be written $X^\mu h_{\mu\nu}=0$. Inserting the expression~\eqref{shortvector} for the superrotation vector field and using radial gauge~\eqref{rg} we find 
\begin{equation}
\rho^3 \tau F_{\tau\rho}  = (\tau\partial_\tau-2)\left[{\rho^3  Y^zh_{\rho z}}-\frac{\rho}{2} \p_z^2Y^zh_{\rho\bar{z}}+{2 \p_zY^z h_{z\bar{z}}}\right]. 
\end{equation}
Under the integral we may integrate by parts with respect to $z$, yielding the expression
\begin{equation}
\label{frt} 
\rho^3 \tau F_{\tau\rho} =Y^z{(\tau\partial_\tau-2)}\left[\rho^3  h_{\rho z} - \frac{\rho}{2}\p_z^2 h_{\rho\bar{z}} - 2\p_z h_{z\bar{z}}\right].
\end{equation}
As in~\cite{Campiglia:2015kxa}, the  boundary conditions are chosen to ensure that the charge is $\tau$-independent and finite for $\rho\to \infty$, so that it does not depend on a choice of slice. Finiteness of the charge requires that the leading $\rho$ behavior is $h_{\rho z}\sim\rho^{-3}, ~h_{z \bar z} \sim \rho^0$, which is compatible with the linearized analysis in the appendix.  Moreover we assume that the Bondi news vanishes at $\csp$. Otherwise, as mentioned above, there are correction terms to the charge~\cite{Wald:1999wa}. The finite and $\tau$-independent final boundary expression for the superrotation charge is
\begin{equation}
\CQ^+(\zeta_Y)=-{1 \over 16\pi} \lim_{\rho\rightarrow\infty}\int_\csp d^2z Y^z \bigg[ \rho^3h_{\rho z}^{(0)} 
- 2\p_z h^{(0)}_{z\bar{z}} \bigg], 
\end{equation}
where the superscript $(0)$ indicates the $\tau$-independent piece of the given metric component.

\subsection{Linearized Bulk Charge}
Having found an expression for the charge $\mathcal{Q}^+$ as a surface integral over $\csp$, we now write a bulk expression for the linearized charge as an integral over $H_3^+$. This involves integrating by parts and using the linearized vacuum Einstein equations.  We denote the linearized charge as $\CQ_S^+(\zeta)$ because, as we shall see, it is the same as the soft part of the full nonlinear charge.  The nonlinearities are incorporated in the next subsection, where we also discuss the validity of the linearized approximation. 

Starting with the boundary definition of the linearized charge $\CQ_S^+(\zeta)$, the desired bulk expression follows from an application of Stokes's theorem and the linearized constraint equations. By construction the bulk charge is the symplectic product of the metric variation $\CL_\zeta g_{\mu\nu}$ produced by $\zeta$ with the linearized metric perturbation $h_{\mu \nu}$,
\begin{equation}
\label{qss} 
\CQ^+_S(\zeta)=\bigl(\CL_\zeta g,h\bigr)_{H^\tau_3}=\int_{H^\tau_3} d\Sigma^\mu P^{\nu \lambda \gamma \sigma}\CL_\zeta g_{\nu \lambda}\overleftrightarrow \nabla_\mu h_{\gamma\sigma},
\end{equation}
where $\bigl(~,~\bigr)_{M_3} $ is the symplectic product on a three-manifold $M_3$,  $H_3^\tau$ is any hyperbolic slice of given $\tau$ and the required components of $P$ (given in full in~\cite{Wald:1999wa}) are given below.  Since the symplectic product is conserved on-shell (assuming appropriate smoothness conditions at $\csp$) this expression does not depend on the choice of hyperbolic slice $\tau$. We will take $\tau\rightarrow\infty$. 
In the quantum theory, $h_{\mu\nu}$ then becomes a free field operator, and commutators with $\CQ^+_S$ formally generate linearized superrotations of the metric on $H_3^+$. 

In the case at hand, the only nonzero component of the metric variation is~\eqref{yex} and we need only the component $P^{zz\zb\zb}={ 1 \over 8\pi \tau^4 \rho^4}$. The linearized charge reduces to the simple expression
\begin{equation}
\label{qs} \CQ^+_S(\zeta_Y)={1 \over 8 \pi}\int_{H^+_3}{d^2zd\rho \over \tau^2 \rho^3}\CL_Y g_{zz}  h^{(0)}_{\bar{z} \bar{z} }=-{1 \over 16 \pi}\int_{H^+_3}{d^2zd\rho \over \rho^3}\p_z^3Y^z h^{(0)}_{\bar{z} \bar{z} },
\end{equation}
where $h^{(0)}_{\bar{z} \bar{z} }$ is the $\tau$-independent part of $h_{\bar{z} \bar{z} }$.

We note that $\CL_Yg_{zz}$, as given in \eqref{yex}, involves only the order $\rho^0$ metric perturbation,\footnote{In Section 5, to facilitate the connection to the soft theorem, a physically equivalent  vector field $\zeta'_w=\zeta_w\left(1+{\cal O}\left({1 \over \rho^2}\right)\right)$ which differs at further subleading orders is introduced. } which has been identified~\cite{Balasubramanian:1999re} as  the holographic stress tensor in the context of AdS$_3$ quantum gravity. This gives a precise connection of  the superrotation generators  for $M_4$ quantum gravity as an uplift of the generator of conformal transformations  for AdS$_3$. More specifically, the soft part of the charge which generates 4D superrotations in the causal domain of $H_3^+$ is the symplectic product on the 3D hyperbolic slice of the linearized 4D metric perturbation with the $Y^z$-variation of the 3D holographic stress tensor.  
 
 \subsection{Exact Bulk Charge}
In the previous subsection, the surface charge $\CQ^+(\zeta_Y)$ on $\csp$ was reexpressed as a bulk integral over 
$H_3^+$ in the linearized approximation. For a generic slice in a generic asymptotically flat spacetime ending on $\csp$, nonlinear corrections are important, and  there is no useful bulk expression for the charge. However, it is natural to take $\tau \to \infty$, in which case (assuming no stable black holes) the slice hugs $\mathcal{I}^+$, the fields become weak, and corrections to the linearized approximation are easily incorporated.  

 In order to obtain the bulk expression on $H_3^+$ from the boundary expression on $\csp$ one integrates by parts and uses the constraint equations
$G_{\tau \mu}=8\pi T_{\tau \mu}$. In the linearized approximation,\footnote{The linearized vacuum equations in hyperbolic coordinates are given in Appendix~\ref{A}.} the nonlinear terms on the left  hand side and the entire right hand side are set to zero. In the full theory, the constraints reduce (for $\tau \to \infty$) to  
\begin{equation}
-16\pi T_{\tau \mu} = -2G_{\tau \mu} = \square h_{\tau\mu}-\nabla_\tau\nabla_\alpha h_\mu{}^\alpha-\nabla^\alpha\nabla_\mu h_{\tau\alpha}+\nabla_\tau\nabla_\mu h + \eta_{\tau\mu}\nabla_\alpha\nabla_\beta h^{\alpha\beta}-\eta_{\tau\mu}\square h, 
\end{equation}
where the stress tensor is understood to contain both matter contributions and the quadratic gravity wave stress tensor. Corrections which are cubic or higher in $h_{\mu\nu}$ vanish for $\tau\to\infty$. 
The full expression for the charge is then
\begin{equation}
\label{rdc}  
\CQ^+(\zeta) =\CQ_S^+(\zeta)+ \CQ_H^+(\zeta),
\end{equation}
where $\CQ_S^+(\zeta)$ is given in~\eqref{qss} and  the hard charge is 
\begin{eqnarray}
\label{hardplus} 
\CQ_H^+(\zeta) & = &  \int_{H_3^+}d\Sigma^\mu\zeta^\nu T_{\mu\nu} \cr
 &=& -\frac{1}{2}\int_{H^+_3}{d^2z\, d\rho\,\rho\,\tau^3}T_{\tau \mu}\zeta^\mu. 
\end{eqnarray}
For $\zeta_Y$ as in~\eqref{shortvector},~\eqref{hardplus} becomes
\begin{eqnarray}
 \CQ_H^+(\zeta_Y) &=& -\frac{1}{2}\int_{H^+_3}{d^2z\,d\rho\,\rho\,\tau^3}\bigg[T_{\tau z}Y^z-\frac{\rho}{2}T_{\tau \rho}\p_zY^z -{1 \over 2\rho^2}T_{\tau \zb}\p_z^2Y^z  \bigg]\cr 
 &=& -\frac{1}{2}\int_{H^+_3}{d^2z\,d\rho\,\rho\,\tau^3}Y^z\bigg[T_{\tau z}+\frac{\rho}{2}\p_z T_{\tau \rho} -{1 \over 2\rho^2}\p_z^2T_{\tau \zb}  \bigg].
\end{eqnarray}
Since the matter stress tensor generates diffeomorphisms on the matter fields, this manifestly generates the hard action of the superrotations. 

\section{Massive Point Particles}
In this section we compute the hard charge for a collection of $N$ massive point particles with inertial trajectories, which are given in Cartesian coordinates by  
\begin{equation} 
x_k^\mu(\lam) = \frac{p_k^\mu}{m_k} \lam + b_k^\mu , 
\end{equation}
where $k=1,\ldots,N$ and $p_k^2=-m_k^2$. We follow the analogous treatment of massless point particles presented in~\cite{Pasterski:2015tva}. The massive point particle trajectories asymptote at late times to a fixed point $(\rho_k,z_k, \zb_k)$ on $H_3^+$ with $\lambda=\tau$. In the coordinates~\eqref{hyperboliccoord} this point is determined by 
\begin{equation} 
\lim_{\tau\rightarrow\infty}\frac{1}{\tau}x_k^\mu(\tau)  ={p_k^\mu\over m_k} = \frac{\rho_k}{2}\left( \begin{array}{c}  1+z_k\bar{z}_k+\rho_k^{-2} \\  z_k+\bar{z}_k \\ -i(z_k-\bar{z}_k) \\ 1-z_k\bar{z}_k-\rho_k^{-2} \end{array} \right). 
\end{equation}
The stress tensor of the $k$th  particle is
\be T_k^{\mu \nu}(X) = \int d\lam \frac{p_k^\mu p_k^\nu}{m_k} \delta^{(4)}(X - x_k(\lam)). \ee
Substituting into the first line of~\eqref{hardplus} we find the simple expression
\begin{equation}
\label{qp}
\CQ_H^+(\zeta)=-\lim_{\lambda\rightarrow\infty}\sum_k (p_{k} \cdot \zeta) |_{x_k^\mu(\lambda)}.
\end{equation}
To easily connect to the soft theorem, we use  the  vector field~\cite{Pasterski:2017kqt, Donnay:2018neh}
\begin{equation}
\label{DPS}
\zeta'_{\mu;w} = \frac{1}{4}\partial_w^3[X^\nu(q_\nu\partial_{\bar{w}}q_\mu - q_\mu\partial_{\bar{w}}q_\nu)\log(-q\cdot X)],
\end{equation}
where $q$ is the null vector that points towards $w$ on $\csp$,
\be q^\mu=(1+w{\bar w},w+{\bar w},-i(w-{\bar w}),1-w{\bar w}).\ee
This vector field satisfies 
\be \zeta'_{w}=\zeta_{w}\left(1+{\cal O}\left({1 \over \rho^2}\right)\right)\ee near  $\csp$ and hence gives the same  total charge as $ \zeta_w$. Since $\mathcal{Q}^{\pm}(\zeta'_w) = \mathcal{Q}^{\pm}(\zeta_w)$, the two vector fields have the same Ward identity and conservation law.\footnote{Their soft and hard parts, however, are not separately equal. We find it curious that, even though their difference is trivial, some computations are easier with $ \zeta'_w$ while others are easier with $ \zeta_w$. } The vector field $\zeta'_w$ arises naturally in the study of conformal primary wavefunctions~\cite{Pasterski:2016qvg,Donnay:2018neh}
as well as in the study of massive matter~\cite{Campiglia:2015kxa}. The utility of $\zeta'_w$ over $\zeta_w$ in the present context is its simple relation to the momentum space version of the subleading soft factor~\cite{Cachazo:2014fwa,Campiglia:2015kxa}. We further define polarization tensors
\be \epsilon^{\mu\nu}_{ww}=\epsilon^{\mu}_w\epsilon^{\nu}_w,~~~ \epsilon^{\mu\nu}_{\bar{w}\bar{w}}=\epsilon^{\mu}_{\bar{w}}\epsilon^{\nu}_{\bar{w}}, ~~~\epsilon^{\mu}_w(w,{\bar w})=\frac{1}{\sqrt{2}}(\bar{w},1,-i,-\bar{w}),~~~\epsilon^{\mu}_{\bar{w}}(w,\bar{w}) = \frac{1}{\sqrt{2}}(w,1,i,-w).\ee
One then finds that~\eqref{qp} becomes, after significant algebra, 
\be \label{hardcharge} \CQ_H^+(\zeta'_w)=\frac{1}{2}\sum_k \int d^2z \frac{1}{w-z} \partial_z^3\left[{p_k^\mu \e_{\mu\nu;\bar{z}\bar{z}}J_k^{\nu\alpha}q_\alpha \over p_k\cdot  q }\right],\ee
where the tensors
\be J_k^{\mu\nu}=x^\mu_kp_k^\nu-x^\nu_kp_k^\mu \ee
are the boost and angular momentum charges of the $k$th particle. The quantity in square brackets in~\eqref{hardcharge} is immediately recognizable as the soft factor in the subleading soft graviton theorem.

\section{Subleading Soft Theorem}
In this section we argue that the Ward identity of our charge implies the subleading soft graviton theorem~\cite{Cachazo:2014fwa, Campiglia:2015kxa}. In~\cite{Kapec:2014opa} the classical conservation law associated to superrotations is expressed as a sum of integrals over $\mathcal{I}^\pm$ in Bondi coordinates,
\begin{equation}
\label{3erf} 
Q_S(Y) + Q_H(Y)=0,
\end{equation}
with 
\begin{eqnarray}
\label{qref}
Q_S (Y)&=& {1 \over 16\pi}\int_{\mathcal{I}^+}d^2zduY^zu\p_z^3N^z_{~\zb}-{1 \over 16\pi}\int_{\mathcal{I}^-}d^2zdvY^zv\p_z^3N^z_{~\zb} \cr
Q_H(Y) & = & \int_{\mathcal{I}^+} d^2z du \, r^2 Y_\zb\left(T_{uz} - \half u\p_zT_{uu}\right)-\int_{\mathcal{I}^-} d^2z dv \, r^2 Y_\zb\left(T_{vz} - \half v\p_zT_{vv}\right),
\end{eqnarray}
where we take $Y^\zb=0$, $N_{zz}$ is the Bondi news, and we raise and lower sphere indices using the round metric on the unit sphere $S^2$. It was shown in~\cite{Kapec:2014opa} that the quantum version of this conservation law is equivalent to the subleading soft graviton theorem~\cite{Cachazo:2014fwa}. This conservation law can be expressed as the equality of two total charges, one incoming and one outgoing. 

In the present paper, in contrast,  we have three hard and three soft charges associated to the three slices $H^+_3$, dS$^0_3,$ and $H^-_3,$ depicted in Figure~\ref{slicing}. We accordingly decompose
\begin{equation}
\label{dxz} 
\begin{aligned} \CQ_S(\zeta'_w) & = \CQ^+_S(\zeta'_w) + \CQ^{0}_S(\zeta'_w) + \CQ^-_S(\zeta'_w) \\
\CQ_H(\zeta'_w) & = \CQ^+_H(\zeta'_w) + \CQ^{0}_H(\zeta'_w) + \CQ^-_H(\zeta'_w). \end{aligned} 
\end{equation}
Here we show $\CQ_S(\zeta'_w) = Q_S({1 \over w-z})$ and $\CQ_H(\zeta'_w) = Q_H({1 \over w-z})$, and therefore that the subleading soft graviton theorem is equivalent to the conservation law on hyperbolic slices
\be \label{3ef} 
\CQ_S + \CQ_H=0.
\end{equation}

 First, we show that 
\begin{equation}
\label{dxcz} 
 \CQ_H(\zeta'_w)=Q_H\left({1 \over w-z}\right). 
\end{equation}
We can consider the hard charge for massive or massless matter. Massive particles cannot reach the asymptotic dS$^0_3$ and therefore contribute only to the $\CQ_H^\pm$ charges. As computed in the previous section, the left hand side is
\begin{equation}
\label{tyP} 
\frac{1}{2}\int d^2z \frac{1}{w-z} \partial_z^3\left[\sum_k {p_k^\mu \e_{\mu\nu;\bar{z}\bar{z}}J_k^{\nu\alpha}q_\alpha \over p_k\cdot  q }-\sum_j {p_j^\mu \e_{\mu\nu;\bar{z}\bar{z}}J_j^{\nu\alpha}q_\alpha \over p_j\cdot  q }\right],
\end{equation}
where $p_k$ are outgoing and $p_j$ are incoming momenta. One finds that the same expression holds when we act with the  hard charge on massless particles, with the momenta $p_k$ taken to be null. This agrees with  $Q_H$ in \eqref{qref} (see~\cite{Kapec:2014opa}) and shows that the hard charges are the same.

Next, we wish to verify agreement between the soft terms evaluated in Bondi and hyperbolic coordinates, i.e.
\begin{equation}
\label{dxpcz}
  Q_S = \CQ^+_S(\zeta'_w) + \CQ^{0}_S(\zeta'_w) + \CQ^-_S(\zeta'_w). 
  \end{equation}
In order to do so, we rewrite the first line in the Bondi expression~\eqref{qref} as 
\begin{equation} 
\label{pl}Q_S=\int_{\mathcal{I}^+} *J+\int_{\mathcal{I}^-} *J, 
\end{equation}
with 
\begin{equation} 
J=P^{\nu \lambda \gamma \sigma}\CL_{\zeta'_w} g_{\nu \lambda} \nabla_\mu h_{\gamma\sigma}dx^\mu. 
\end{equation}
Note the use here of $\nabla_\mu$ rather than $\overleftrightarrow{\nabla}_\mu$, which appears in the gravitational symplectic pairing~\eqref{qss}. Since $\int du\,u\,N_{\zb \zb}$ is  the subleading soft graviton insertion, and the Bondi news, up to superrotations, falls off faster than $1\over u$ (or $1 \over v$) at the boundaries of $\mathcal{I}$~\cite{Christodoulou:1993uv,Kapec:2014opa}, we do not expect new soft contributions from ``capping" $\mathcal{I}^\pm$ at past and future timelike infinity $i^\pm$. The soft charge~\eqref{pl} then becomes
\begin{equation}
Q_S = \int_{H_3^+ \cup {\rm{dS}}^0_3 \cup H_3^-} *J. 
\end{equation}
Now that we are integrating over a surface without boundary, we are free to switch from $\nabla_\mu$ to $\half\overleftrightarrow{\nabla}_\mu$ because they differ by an exact form. The resulting integrand is the same one defining our soft charges, so we have
\begin{equation}
Q_S = \CQ^+_S(\zeta'_w) + \CQ^{0}_S(\zeta'_w) + \CQ^-_S(\zeta'_w). 
\end{equation}
Since it has already been shown that the quantum version of~\eqref{3erf} is the subleading soft graviton theorem, we have demonstrated the desired equivalence of the quantum matrix elements of $\CQ_S(\zeta'_w) + \CQ_H(\zeta'_w) = 0$ to the subleading soft graviton theorem. 

\section{Celestial Stress Tensor}
So far we have not explicitly shown that the action of the charge $\CQ_S(\zeta'_w)$, as suggested by the form of~\eqref{yex}, corresponds to conformal transformations on the celestial sphere.  A fast way to do this is to expand the  Bondi news in asymptotic graviton creation and annihilation operators and then use the results of~\cite{Kapec:2016jld}. One finds \begin{equation}
\label{stress}i\CQ_S(\zeta'_w)= {\cal T}^{KMRS}_{ww},
\end{equation}
where  $ {\cal T}^{KMRS}_{ww}$ is the subleading soft graviton mode~\cite{Kapec:2016jld} 
\begin{equation}
{\cal T}^{KMRS}_{ww}\equiv \frac{3}{\pi\sqrt{32\pi G}}\lim_{\omega\to 0}(1+\omega\p_\omega)\int{d^2z\over (w-z)^4}\big(a_-(\omega q)-a_+^\dagger(\omega q)\big),
\end{equation}
and $a_-$ and $a_+^\dagger$ are asymptotic graviton annihilation and creation operators. As shown in~\cite{Kapec:2016jld}, by reverse-engineering the subleading soft theorem of~\cite{Cachazo:2014fwa},  normal-ordered insertions of $  {\cal T}^{KMRS}_{ww}$ in the 4D $\cal S$-matrix obey the Ward identities of a 2D stress tensor, and therefore generate conformal transformations of the celestial sphere. In particular, if we pick a contour $C$ and integrate
\begin{equation}
\frac{1}{2\pi i }\oint_C dwY^w {\cal T}_{ww}^{KMRS}
\end{equation}
for an arbitrary $Y^w(w)$, the corresponding $\cal S$-matrix insertions generate conformal transformations on the celestial sphere associated to the holomorphic extension of $Y^w$ into the interior of $C$. Thus $i\CQ_S(\zeta'_w)$ is the celestial stress tensor. 

\section{Dual Stress Tensor}
In $U(1)$ gauge theory, large electric gauge transformations $\delta_\e$ on the celestial sphere are generated by a current $J_w$ with left/right conformal dimensions $(1,0)$~\cite{Kapec:2015ena,Nande:2017dba,Donnay:2018neh}. This current can be constructed from the symplectic product of the Goldstone mode wavefunction 
$\delta_\e A_\mu$ with the linearized gauge field operator at null infinity. The Goldstone wavefunction has a symplectic partner which is not pure gauge and leads to a second, symplectically conjugate  $(1,0)$ current $S_w$~\cite{Nande:2017dba}. $S_w$ is related to large magnetic gauge transformations~\cite{Strominger:2015bla}.

We note briefly here that a  similar structure exists for the stress tensor ${\cal T}^{KMRS}_{ww}$, which, like $J_w$, is constructed from the symplectic product with a $(2,0)$ Goldstone mode wavefunction $\delta_Yg_{\mu\nu}$.  In the normalization conventions of~\cite{Donnay:2018neh}, to which we refer the reader for  details, the  $(2,0)$ Goldstone mode is\footnote{In~\cite{Donnay:2018neh} this mode is denoted  $\widetilde{h^{\Delta=0}}_{\mu \nu;ww}$, where  the tilde indicates the fact that it is the shadow of a mode with conformal weight 0 in the basis~\eqref{cpw}.}
 \begin{equation}
h^{\rm Goldstone}_{\mu \nu; ww} =-\frac{1}{6}\left[ \nabla_{\mu} \zeta'_{\nu;w} + \nabla_{\nu} \zeta'_{\mu;w}\right].
\end{equation}
This wavefunction  has a $(2,0)$ symplectic partner that is not pure gauge. The symplectic partner is the $\Delta = 2$ conformal primary wavefunction~\cite{Donnay:2018neh}, where for general $\Delta$
\begin{equation}
\label{cpw}h_{\mu \nu; ww }^{\Delta,\pm}(X^\mu; w, \bar{w}) = \half \frac{[(-q \cdot X) \p_w q_\mu +(\p_\mu q \cdot X) q_\mu] [(-q \cdot X) \p_w q_\nu + (\p_w q \cdot X) q_\nu]}{(-q \cdot X \mp i\eps)^{\Delta+2}}. 
\end{equation}
These solutions are labelled by  $\pm$ for ingoing versus outgoing,  the complex parameter $w$ for the point where the radiation flux crosses the celestial sphere, and $\Delta$ for the $SL(2,\mathbb{C})$ conformal weight.
In hyperbolic coordinates $(\tau, \rho, z, \zb)$\footnote{We note that these modes generically have radiation flux though $\mathcal{CS}^+$~\cite{Donnay:2018neh} and so 
do not obey the boundary conditions for the charge defined on that surface.}
\begin{equation}
h_{\mu \nu;ww}^{\Delta,\pm} = \frac{\tau^{2-\Delta}}{2(|w-z|^2 + \rho^{-2} \mp i\eps)^{\Delta+2}} \left( \begin{array}{cccc} 0 & 0 & 0 & 0 \\
0 & \frac{4 (\bar{w} - \zb)^2}{\rho^{\Delta+4}} & \frac{2(\bar{w}-\zb)^3}{\rho^{\Delta+1}} & \frac{-2(\bar{w} - \zb)}{\rho^{\Delta+3}} \\
0 & \frac{2(\bar{w} - \zb)^3}{\rho^{\Delta+1}} & \frac{(\bar{w} - \zb)^4}{\rho^{\Delta-2}} & \frac{-(\bar{w} - \zb)^2}{\rho^\Delta} \\
0 & \frac{-2(\bar{w} - \zb)}{\rho^{\Delta+3}} & \frac{-(\bar{w} - \zb)^2}{\rho^\Delta} & \frac{1}{\rho^{\Delta+2}} \end{array} \right). 
\end{equation}
For $\Delta= 2$ one finds the simple result~\cite{Pasterski:2017kqt,Donnay:2018neh}
\be \label{deltatwo} h^{2}_{\mu \nu; ww}={1 \over \tau^2}h^{\rm Goldstone}_{\mu \nu; ww},\ee
which is not  a pure diffeomorphism. The symplectic product  \eqref{qss} of these two modes on $H_3^+$  is\footnote{Useful formulae for evaluating these integrals can be found  in~\cite{Dolan:2003hv,SimmonsDuffin:2012uy,Pasterski:2017kqt}. }
\begin{equation}
\label{Qsoft}
\bigl( h^{\rm Goldstone}_{ ww},h_{vv}^{2+i\varepsilon}\bigr)_{H_3^+}=\frac{\pi}{48(w-v)^4}\delta(\varepsilon).
\end{equation}
This resembles an off-diagonal central charge. 
The symplectic product  over a complete spacelike Cauchy  slice $\Sigma_3$ is  
\begin{equation}\label{css}\bigl( h^{\rm Goldstone}_{ ww},h_{vv}^{2+i\varepsilon}\bigr)_{\Sigma_3} =  -\varepsilon\frac{i\pi^2}{6(w-v)^4}\delta(\varepsilon).
\end{equation}
Na\"ively, the right hand side vanishes due to the factor of the imaginary part of the conformal weight $\varepsilon$. However, we leave it in this form as in some contexts  there may be compensating conformally soft poles in  $\varepsilon$.
A second conformal weight (2,0) operator on the celestial sphere (in addition to ${\cal T}^{KMRS}_{ww}$) can be constructed explicitly from the mode~\eqref{deltatwo}. Potential implications of  two weight (2,0) operators for the structure of the soft gravitational ${\cal S}$-matrix are left to future work. 

\section*{Acknowledgements}
This work was supported in part by NSF grants 1205550, 1745303, and 1144152, the John Templeton Foundation, and the Hertz Foundation.   We are grateful to Laura Donnay, Dan Kapec, Monica Pate, Andrea Puhm, Ana-Maria Raclariu, and Shu-Heng Shao for useful discussions. 

 \appendix

\section{Linearized Einstein Equations}
\label{A}
In radial gauge, $h_{\tau \mu} = 0$, the Einstein equations take the form
  \begin{eqnarray}
  \label{einstein}
G_{\tau\tau} &=& \frac{\rho^2}{\tau^4}\left( \tau\p_\tau+\rho\p_\rho+3-\frac{2}{\rho^2}\p_z\p_\bz\right)h_{\rho\rho}+\frac{2}{\rho\tau^4}(\rho\p_\rho+2)(\p_\bz h_{\rho z}+\p_z h_{\rho\bz})\cr
& & +\frac{2}{\rho^4\tau^4}(\p_\bz^2h_{zz}+\p_z^2h_{\bz\bz}) +  \frac{2}{\rho^2\tau^4}\left(2\tau\p_\tau-\rho^2\p_\rho^2-\frac{2}{\rho^2}\p_z\p_\bz-2\right)h_{z\bz}\cr
 G_{\tau\rho}&= &(\tau\p_\tau-2)\left[\frac{\rho}{\tau^3}h_{\rho\rho}+\frac{1}{\rho^2\tau^3}(\p_\bz h_{\rho z}+\p_z h_{\rho\bz})+\frac{2}{\rho^3\tau^3}(1-\rho\partial_\rho)h_{z\bz}\right]\cr
 G_{\rho\rho}&=&-\frac{1}{\tau^2}h_{\rho\rho}-\frac{2}{\rho^3\tau^2}(\p_\bz h_{\rho z}+\p_z h_{\rho\bz})-\frac{2}{\rho^6\tau^2}(\p_\bz^2h_{zz}+\p_z^2 h_{\bz\bz})\cr
 & & -\frac{2}{\rho^4\tau^2}\left(\tau^2\p_\tau^2-\tau\p_\tau-\rho\p_\rho+2-\frac{2}{\rho^2}\p_z\p_\bz\right)h_{z\bz}\cr
 G_{\tau z}&=&(\tau\p_\tau-2)\left[-\frac{\rho^2}{2\tau^3}\p_z h_{\rho\rho}   +\frac{\rho}{2\tau^3}(\rho\p_\rho+3)h_{\rho z}+\frac{1}{\rho^2\tau^3} (\p_\bz h_{zz}-\p_z h_{z\bz})\right]\cr
G_{\rho z}&=&\frac{\rho}{2\tau^2}\p_z h_{\rho\rho}+\frac{1}{2\tau^2}\left(\tau^2\p_\tau^2-\tau\p_\tau-\frac{2}{\rho^2}\p_z\p_\bz\right)h_{\rho z}+\frac{1}{\rho^2\tau^2}\p_z^2h_{\rho\bz}\cr
& & +\frac{1}{\rho^3\tau^2}(\rho\p_\rho-2)(\p_\bz h_{zz}-\p_z h_{z\bz})\cr
G_{zz}&=&-\frac{\rho^2}{2\tau^2}\p_z^2 h_{\rho\rho}+\frac{\rho}{\tau^2}(\rho\p_\rho+1)\p_zh_{\rho z}+\frac{1}{2\tau^2}(\tau^2\p_\tau^2-\tau\p_\tau-\rho^2\p_\rho^2+\rho\p_\rho)h_{zz}\cr
G_{z\bz}&=& \frac{\rho^4}{4\tau^2}\left(-\tau^2\p_\tau^2+\tau\p_\tau -\rho\p_\rho-4+\frac{2}{\rho^2}\p_z\p_\bz\right)h_{\rho\rho} +\frac{1}{2\tau^2}(-\tau^2\p_\tau^2 +\tau\p_\tau+\rho^2\p_\rho^2-\rho\p_\rho)h_{z\bz}\cr
&  & -  \frac{\rho}{2\tau^2}(\rho\p_\rho+1)(\p_\bz h_{\rho z}+\p_z h_{\rho \bz}).
\end{eqnarray}
Note that the Einstein equations completely decouple under different $\tau$ scalings, so it is natural to decompose the metric in a $\tau$ expansion as $h_{\mu\nu} = \sum_{n} \tau^{-n}h_{\mu\nu}^{(n)}(\rho,z,\bar{z})$. 

Working in ``on-shell gauge" of the free Einstein equations we arrived at an equation for $h_{zz}^{(0)}$ by itself. The gauge assumes that
\be \label{radial} X^\mu h_{\mu \nu} = 0 \ee
\be \nabla^\mu h_{\mu \nu} = 0 \ee
\be g^{\mu \nu} h_{\mu \nu} = 0, \ee
where $X^\mu$ are Cartesian coordinates. Note that in hyperbolic coordinates~\eqref{radial} is equivalent to $h_{\tau \mu} = 0$. The $G_{\tau \mu}$ equations all follow from these gauge conditions, and $G_{\rho \rho}$ and $G_{z\zb}$ are equivalent in this gauge. The $G_{zz}$ equation can be used to eliminate $h_{\rho z}^{(0)}$ in favor of $h_{zz}^{(0)}$ (up to integration constants). Plugging into $G_{\rho z}$ gives
\be \label{recursive} 0 = \rho^4 (\rho \partial_\rho + 4)(\rho \partial_\rho + 2)(\rho \partial_\rho - 2) \rho \partial_\rho h_{zz}^{(0)} + 8\rho^2 (\rho \partial_\rho + 2)(\rho \partial_\rho - 2) \partial_z \partial_\zb h_{zz}^{(0)} + 16 (\partial_z \partial_\zb)^2 h_{zz}^{(0)}. \ee
Given a solution of~\eqref{recursive}, the other metric components in the gauge~\eqref{radial} are constrained. 

Linearized metric perturbations along a vector field $\xi^\mu \p_\mu$ are given by
\begin{eqnarray}
\delta g_{\tau\tau}&= & \frac{2}{\tau} \tau\p_\tau \xi_\tau \cr
\delta g_{\tau\rho} &=  &\p_\rho \xi_\tau + \frac{1}{\tau}(\tau\p_\tau -2)\xi_\rho \cr
\delta g_{\rho\rho}&=& 2(\p_\rho +\frac{1}{\rho})\xi_\rho-\frac{2\tau}{\rho^2}\xi_\tau \cr
\delta g_{\tau z}&= & \frac{1}{\tau}(\tau\p_\tau-2) \xi_z + \p_z \xi_\tau  \cr
\delta g_{\rho z} &=&(\p_\rho-\frac{2}{\rho}) \xi_z + \p_z \xi_\rho\cr
\delta g_{zz} &= & 2\p_z \xi_z\cr
\delta g_{z\bz} &= &\p_z \xi_\bz+\p_\bz \xi_z+\rho^3\xi_\rho-\rho^2\tau\xi_\tau.
\end{eqnarray}
Setting $\delta g_{\tau \mu} = 0$, 
we must have $\p_\tau \xi_\tau = 0$. We can satisfy the conditions with
$\xi_\tau = 0$ and $\xi_\rho, \xi_z, \xi_\bz \propto \tau^2$,
but this is not completely general. We can also let $\xi_\tau(\rho, z, \bz)$ be a generic function and choose the $\mathcal{O}(\tau)$ pieces of the other components accordingly. The general solution, using $\tau$ weight notation $\xi_\mu^{(n)}$, is
\begin{eqnarray} 
\xi_\tau & = & \xi_\tau^{(0)}(\rho, z, \bz) \cr
\xi_\rho & =  &\tau^2 \xi_\rho^{(-2)}(\rho, z, \bz) + \tau \p_\rho \xi_\tau^{(0)} \cr
\xi_z & = &\tau^2 \xi_z^{(-2)}(\rho, z, \bz) + \tau \p_z \xi_\tau^{(0)} \cr
\xi_\bz & = & \tau^2 \xi_\bz^{(-2)}(\rho, z, \bz) + \tau \p_\bz \xi_\tau^{(0)}. 
\end{eqnarray}
Here we treat the $\tau$ dependence as not included in $\xi_\mu^{(n)}$. We see the free data for these residual diffeomorphisms are four free functions of three variables, and that these free functions only affect the $h_{\mu\nu}^{(-1)}$ and $h_{\mu\nu}^{(-2)}$ pieces of the metric in hyperbolic coordinates.

\bibliography{flatspace}
\bibliographystyle{utphys}
\end{document}